\documentstyle[11pt,twoside,jltp]{article}
\input{epsf.tex}

\title{Interstitial He and Ne in Nanotube Bundles}

\author{George Stan$^*$, Vincent H. Crespi$^*$, Milton W. Cole$^*$\\
and Massimo Boninsegni$^{\dag}$}

\address{
$^*$Department of Physics and Center for Materials Physics,\\
 The Pennsylvania State University,\\ 
104 Davey Lab, University Park, PA 16802, USA\\
$^{\dag}$Department of Physics, San Diego State University,\\ 
San Diego, CA 92182, USA}

\runninghead{Stan, Crespi, Cole and Boninsegni}{Interstitial He and Ne in 
Nanotube Bundles}
       
\begin{document}

\begin{abstract}
We explore the properties of atoms confined to the interstitial regions
within a carbon nanotube bundle. We find that He and Ne atoms are of ideal
size for physisorption interactions, so that their binding energies are
much greater there than on planar surfaces of any known material. Hence
high density phases exist at even small vapor pressure. There can result
extraordinary anisotropic liquids or crystalline phases, depending on the
magnitude of the corrugation within the interstitial channels.

\end{abstract}

\maketitle

\vspace{0.15in}

The bulk synthesis of carbon nanotubes has led to several
investigations of molecular sorption within these tubes. Tubes readily
imbibe those materials with low liquid-vapor surface
tensions\cite{iijima}, in agreement with a qualitative argument based
on the ratio of the cohesive energy to the adhesive energy.  One might
expect, therefore, that inert gases would be strong absorbers within
these tubes; this belief was confirmed in both recent calculations and
experiments\cite{exp,calc}. Here we explore adsorption in the linear
interstitial channels (IC's) between close-packed nanotubes within a
bundle or rope. Small atoms, such as He and Ne, fit well in these
IC's. Since this environment has a large number of nearby C
atoms at an optimal distance, the binding is stronger than in
any other known charge-neutral system.
\begin{figure}[ht]
\label {fig:pot}
\vskip 7.8 cm
\includegraphics{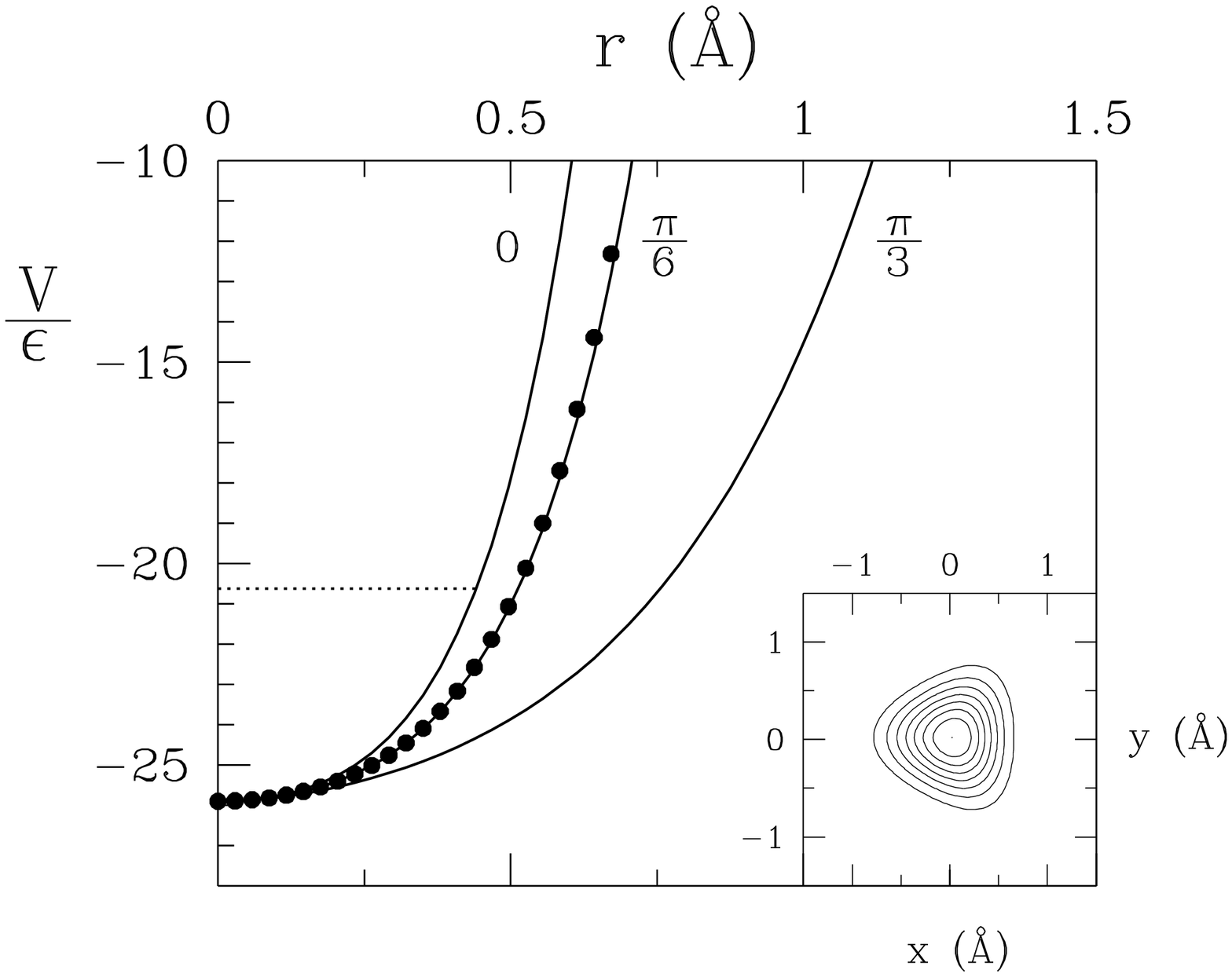}
\small {\noindent
Fig. 1. Axially averaged potential energy for a helium atom as a function of
distance from the channel center for three azimuthal angles relative
to the direction of a neighboring tube center. Also shown are the
angularly averaged potential (circles) and the ground state energy in
this potential (dotted line). The inset shows the ground state
wavefunction within the axially averaged (but not angularly averaged)
potential; the binding energy is within $2-4$ K of the angularly
averaged result. Contour lines are equally spaced in units of
1/8 of the central value.}
\end{figure}
 The present paper describes our initial model  
calculations which establish this fact. After considering the energetics of 
individual atoms, we describe a variety of unique collective phenomena 
which may occur.

Using the pairwise summation of atomic interactions with C,
we have computed the potential energies of He and Ne within the IC's
(which we assume to be infinitely long and perfect); $\epsilon$ is the well 
depth of the pair potential ($16.2$ K for He and $32.1$ K for Ne). 
The study has been done as a function of separation $d$ between the axes 
of adjacent channels. Fig. 1 shows the resulting potential for the experimental
 separation \cite{smalley}, $d=9.8$ \AA\ (tube radius 6.9 \AA). 
Fig. 2 reveals that this value of $d$ provides nearly the greatest possible 
attraction for both Ne and He atoms; the potential on the IC axis is 
approximately a factor of two greater than the well depth within a tube of the
 same radius and is $\sim$150\% greater than the well depth on a graphite 
surface. 
Note that graphite provides the most attractive potential of any 
planar surface known. These comparisons imply that the uptake of these gases 
within the IC's ought to be significant and easily measured. We may compute the
 sorption using a variety of alternative models, several of which we discuss 
here.
\begin{figure}[ht]
\label{fig:gr_state}     
\vskip 7.8 cm
\includegraphics{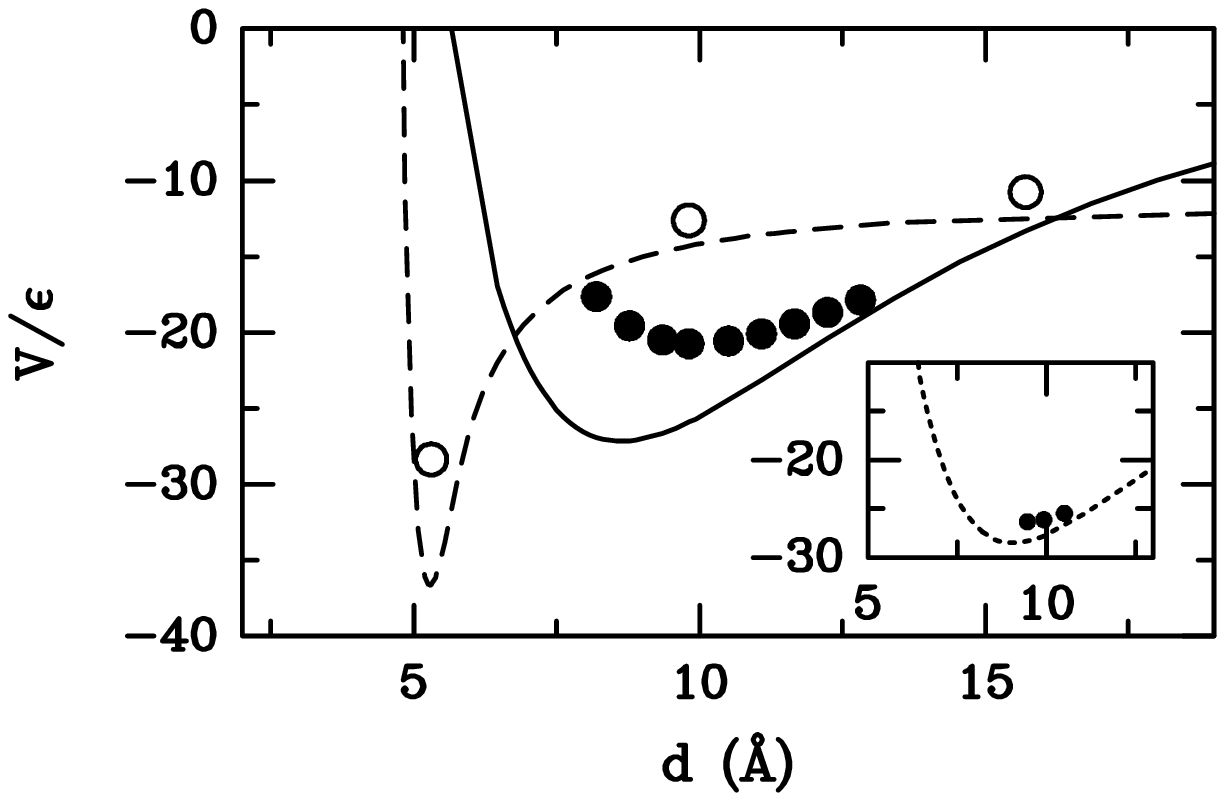}
\small {\noindent
Fig. 2. Minimum of the potential energy for He (solid line) as a function of 
IC separation $d$, as well as the ground state energy (filled circles). Shown 
for comparison is the analogous result for  He adsorption within a tube 
(dashed line and open circles). The case of Ne appears in the inset. Typically,
 $d \sim 10$\AA.}
\end{figure}

One possible assumption is to neglect the corrugation of the potential
along the $z$ axis, i.e. the axial direction of the IC's, and to angularly average the transverse potential.  We  then solve the 
single particle Schr\"{o}dinger equation in the cylindrically symmetric 
potential; the potential and the lowest eigenfunction 
for He are shown in Fig. 1. The ground state energies for He and Ne 
appear in Fig. 2.
For He, the computed binding energy at $d=9.8$ \AA\ is a 
factor 2.3 greater than the experimental value (144 K) on graphite
\cite{elgin}. 
For Ne, the analogous ratio is also 2.3 (because the distance scale of the 
interaction  is similar to that of He) \cite{constabaris,steele}. In order 
to make quantitative predictions about the sorption in the IC's, we present 
here several simplified model calculations. A more thorough investigation 
is in progress. We believe that the finding of significant sorption will 
not be affected by refinements of the calculations.

At sufficiently low density (and not too low $T$), we assume that
the adsorbate is an ideal classical 1d gas; hence, the resulting chemical potential
satisfies $\mu=E_\perp + k_B T\ ln( \rho \lambda)$. Here 
$\rho$ is the 1d density within an IC, $E_{\perp}$ is the transverse 
ground state energy and $\lambda$ is the de Broglie thermal wavelength. From 
this relation and the analogous expression for the coexisting vapor (of density
 $n$ and pressure $P$), we find that $\rho = n \lambda^2 \ {\rm exp} 
(-E_\perp /k_B T$). 
This relation implies for He that an extraordinarily low pressure,
$P \simeq 10^{-34}$ atm, yields a 1d gas with spacing 5 \AA\ at $T=4.2$ K.
\begin{figure}[ht]
\label{fig:eq_state}
\vskip 8.0 cm
\includegraphics{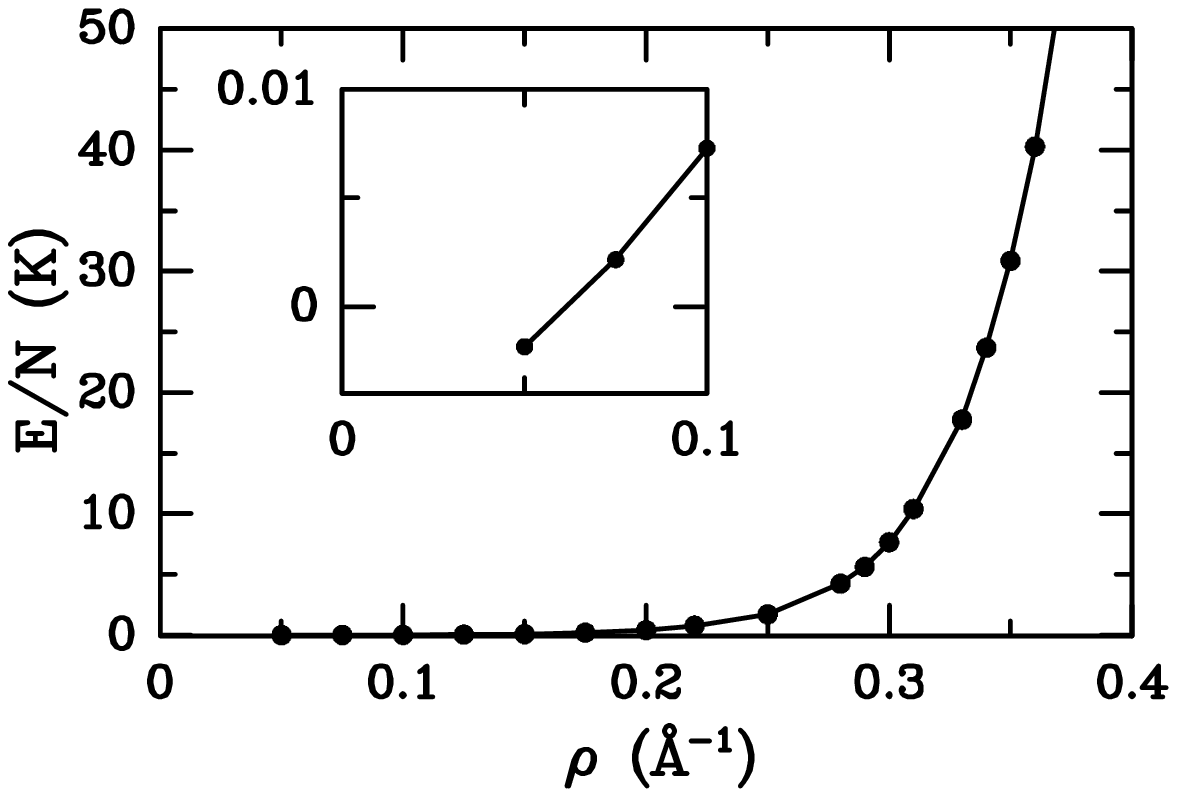}
\small {\noindent
Fig. 3. The energy per particle of 1d He as a function of density at $T=0$. The inset shows the low density regime.}
\end{figure}
 One requires a pressure 20 orders of magnitude higher to achieve the same interparticle spacing on a graphite surface!

Quantum and interaction effects enhance interest in this problem.
Hence we have addressed the extreme case of 1d He at $T=0$.   
We have determined the equation of state  of this system numerically, 
using the Diffusion Monte Carlo method with the Aziz interatomic potential 
\cite{aziz}. Fig.3 shows our results. While such a 1d system undergoes no finite temperature transition, we have explored the effects of interactions between atoms in neighboring IC's. Using a variational approach we have demonstrated the existence of  a self-bound liquid at a density around 0.05 \AA$^{-1}$, with a binding energy per particle (and transition temperature) $\simeq$ tens of mK. We believe that it is a remarkable superfluid without Bose-Einstein condensation (as found also in 2d He at low $T$ \cite{kosterlitz}).

 From the data in Fig. 3, we find that the chemical potential of 
the 1d system increases rapidly above a density 0.25 \AA$^{-1}$ (rising from 
$\sim$50 K to over 300 K at 0.35 \AA$^{-1}$) due to both zero-point 
energy and the repulsive interactions. 
At density $<$ 0.25 \AA$^{-1}$, however, the 
net chemical potential, including the IC single particle energy $E_\perp$, 
lies below the single particle chemical potential on graphite. Hence we 
predict that a high density interstitial fluid would 
form at low $\mu$ in the IC's before any significant gas would adsorb on a graphite surface at low $T$. 
\begin{figure}[ht]
\label{fig:energy}
\vskip 7.8 cm
\includegraphics{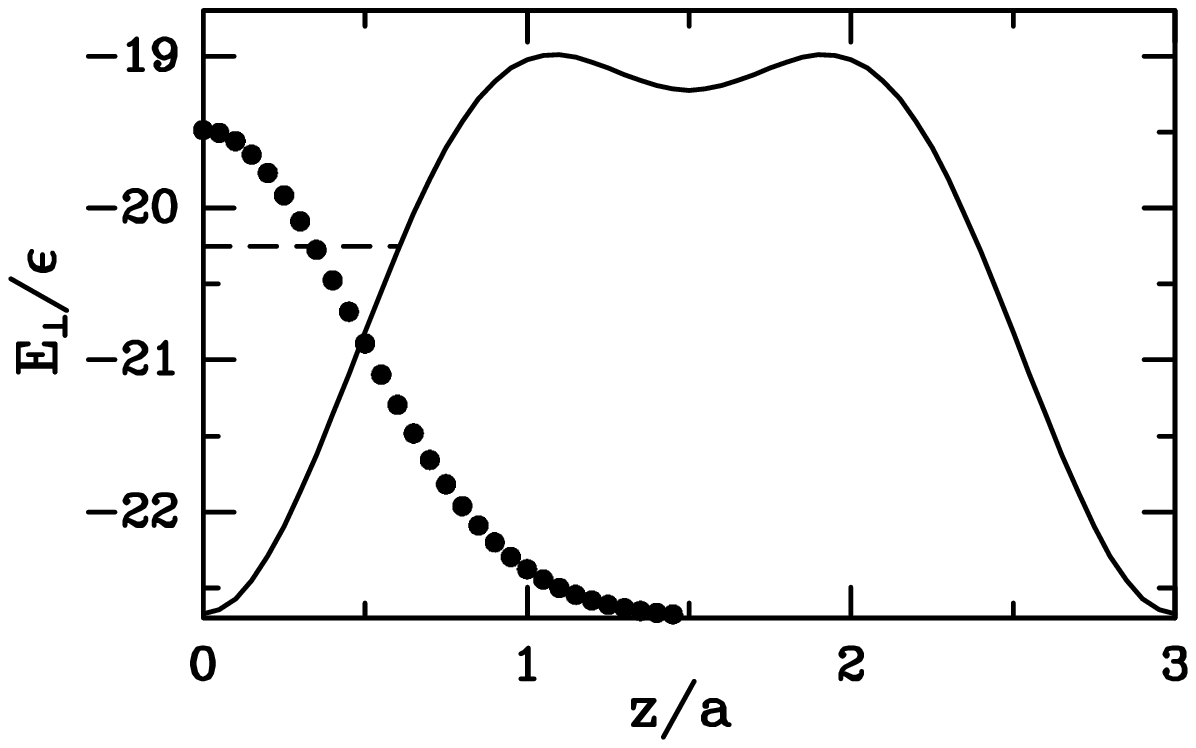}
\small {\noindent
Fig. 4. The transverse eigenvalue as a function of axial position (solid line)
 and the localized longitudinal wave function (unnormalized, circles) and 
eigenvalue (dashed line) are shown for the corrugated channel. $a=1.415$ \AA\ 
is the C-C bond length.}
\end{figure}

We next examine the assumption that the atomic motion along the channels is
that of a free particle. The corrugation of the potential on a
graphite surface is known to be a $\sim$15\% effect, leading to band 
mass $m^* \sim 1.05\ m$ \cite{guo}. 
In the very confined environment of the IC, the energy scale associated to 
corrugation is now magnified so that band structure effects are 
considerable.
This problem merits a fully three dimensional calculation, which
we have not yet performed. Rather, we have solved a 2d Schr\"{o}dinger equation
 for a series of fixed positions $z$ along the IC axis; the resulting 
eigenvalue $E_\perp$ becomes an input into a 1d band structure calculation 
\cite{cole}. We observe in Fig. 4 that this effective potential corresponds to 
a highly localized system.  This property is manifested in the energy level 
and wave function shown, which are computed with a simplified harmonic 
approximation. The system of many atoms moving in such a potential has 
remarkable properties. The atoms may be so strongly localized that the system realizes the (usually artificial) lattice gas model. The ground state would then be a unique quantum crystal which can be characterized in a diffraction, 
thermodynamic, or NMR experiment. We shall report on its properties in a future
 publication.

The corrugations treated here are extrema within the available possibilities.
Uniform aligned tubes yield the maximum possible corrugation. We are currently
 evaluating alternative geometries to see whether the intriguing fluid, derived
 with the smooth potential model, may also be realized by appropriate choice of
 geometry.

We summarize our results as follows. Calculations indicate that He and Ne
are well suited to squeeze into the interstitial channels between C
nanotubes. The binding there is so strong that a high density fluid is
reached at vapor pressures so low that negligible adsorption occurs on
basal plane graphite. When cooperative effects between atoms in neighboring
tubes are included, we find exotic anisotropic condensed phases to occur.
These are being investigated currently.

\section*{ACKNOWLEDGMENTS}
This research has been supported by the National Science Foundation under 
research grants DMR-9623961 and DMR-9802803, by the San Diego State University 
Foundation and the Petroleum Research Foundation, and has benefited from
discussion with L. W. Bruch, Yong-Baek Kim, Moses Chan, Karl Johnson, Paul 
Lammert, Tom Gramila, Jayanth Banavar and Paul Sokol.

\end{document}